\documentclass[letterpaper,twocolumn,
prl
,superscriptaddress,email
]{revtex4}

\usepackage{soul}
\usepackage[pdftex]{graphicx}
\usepackage{amssymb}
\usepackage{mathrsfs}
\usepackage{amsmath,amsfonts,latexsym}
\usepackage{array,tabularx,xcolor}
\usepackage{dcolumn} 
\usepackage[normalem]{ulem}
\usepackage{comment}
\usepackage{bm}
\usepackage{doi}
\usepackage{hyperref}

\hypersetup{
   colorlinks=true,
   linkcolor=magenta,
   citecolor=magenta,
   filecolor=magenta,
   urlcolor=red,
}

\newcommand{\ket}[1]{| #1 \rangle}
\newcommand{\bra}[1]{\langle #1 |}

\begin{document}

\title{Overcomplete quantum tomography of a path-entangled two-photon state}

\author{L. De Santis}
\affiliation{Centre de Nanosciences et de Nanotechnologies, CNRS, Univ. Paris-Sud, Universit\'e Paris-Saclay, C2N -- Marcoussis, 91460 Marcoussis, France}

\author{G. Coppola}
\affiliation{Centre de Nanosciences et de Nanotechnologies, CNRS, Univ. Paris-Sud, Universit\'e Paris-Saclay, C2N -- Marcoussis, 91460 Marcoussis, France}

\author{C. Ant\'on}
\affiliation{Centre de Nanosciences et de Nanotechnologies, CNRS, Univ. Paris-Sud, Universit\'e Paris-Saclay, C2N -- Marcoussis, 91460 Marcoussis, France}

\author{N. Somaschi}
\affiliation{Centre de Nanosciences et de Nanotechnologies, CNRS, Univ. Paris-Sud, Universit\'e Paris-Saclay, C2N -- Marcoussis, 91460 Marcoussis, France}

\author{C. G\'omez}
\affiliation{Centre de Nanosciences et de Nanotechnologies, CNRS, Univ. Paris-Sud, Universit\'e Paris-Saclay, C2N -- Marcoussis, 91460 Marcoussis, France}

\author{A. Lema\^itre}
\affiliation{Centre de Nanosciences et de Nanotechnologies, CNRS, Univ. Paris-Sud, Universit\'e Paris-Saclay, C2N -- Marcoussis, 91460 Marcoussis, France}

\author{I. Sagnes}
\affiliation{Centre de Nanosciences et de Nanotechnologies, CNRS, Univ. Paris-Sud, Universit\'e Paris-Saclay, C2N -- Marcoussis, 91460 Marcoussis, France}

\author{L. Lanco}
\affiliation{Centre de Nanosciences et de Nanotechnologies, CNRS, Univ. Paris-Sud, Universit\'e Paris-Saclay, C2N -- Marcoussis, 91460 Marcoussis, France}
\affiliation{Universit\'e Paris Diderot -- Paris 7, 75205 Paris CEDEX 13, France}

\author{J. C. Loredo}
\affiliation{Centre de Nanosciences et de Nanotechnologies, CNRS, Univ. Paris-Sud, Universit\'e Paris-Saclay, C2N -- Marcoussis, 91460 Marcoussis, France}

\author{O. Krebs}
\email{olivier.krebs@c2n.upsaclay.fr}

\affiliation{Centre de Nanosciences et de Nanotechnologies, CNRS, Univ. Paris-Sud, Universit\'e Paris-Saclay, C2N -- Marcoussis, 91460 Marcoussis, France}

\author{P. Senellart}
\email{pascale.senellart-mardon@c2n.upsaclay.fr}
\affiliation{Centre de Nanosciences et de Nanotechnologies, CNRS, Univ. Paris-Sud, Universit\'e Paris-Saclay, C2N -- Marcoussis, 91460 Marcoussis, France}

\date{\today}

\begin{abstract}
Path-entangled N-photon states can be obtained through the coalescence of indistinguishable photons inside linear networks. They are  key resources for quantum enhanced metrology, quantum imaging, as well as quantum computation based on quantum walks. However, the quantum tomography of path-entangled indistinguishable photons is still in its infancy as it requires multiple  phase estimations increasing rapidly with N. Here, we propose and implement a method to measure the quantum tomography of path-entangled two-photon states. A two-photon state is generated through the Hong-Ou-Mandel interference of highly indistinguishable single photons emitted by a semiconductor quantum dot-cavity device. To access both the populations and the coherences of the path-encoded density matrix, we introduce an ancilla spatial mode and perform  photon correlations as a function of a single phase in a split Mach-Zehnder interferometer. We discuss the accuracy of standard quantum tomography techniques and  show that an overcomplete data set can reveal spatial coherences  that could be otherwise hidden due to limited or noisy statistics. Finally, we extend our analysis to extract the truly   indistinguishable  part of the density matrix, which allows us to identify the main origin for the imperfect fidelity to the maximally entangled state.
\end{abstract}

\maketitle



\noindent{\bf Introduction}

Path-entanglement is an important resource in the field of precision measurements, where the use of  entangled particles provides accuracy beyond the  standard quantum limit. A textbook example is the quantum enhanced optical phase measurement~\cite{Nagata726}, that has already shown  important applications in the field of microscopy~\cite{micro1,micro2,micro3}, lithography~\cite{litho1,litho2}, biology sensing~\cite{bio1,mitchell2012} as well as gravitational-wave detection~\cite{Dowling}. The quantum advantage arises from the use of path entanglement in interferometric protocols.
For instance, a  path-entangled N-photon state in the form of  $\ket{N0}+e^{i\phi}\ket{0N}$, referred to as a N00N state, enables an N-fold enhancement in the phase resolution with a measurement sensitivity of $\Delta \phi=\frac{1}{N}$, beyond the standard quantum limit of $\Delta \phi=\frac{1}{\sqrt{N}}$~\cite{phase1}. Path-entanglement has also been proposed as a resource for quantum computing, both for intermediate---i.e., non-universal---tasks like Boson sampling\cite{aaronson}, as well as for universal quantum computation using quantum walks of indistinguishable particles~\cite{QWcomputing1,QWcomputing2,QWcomputing3}.

Various schemes are proposed to generate N00N states using  beam-splitters, ancillary photons and post-selection  for path-entangled states~\cite{WhiteN00N2003,DowlingGeneration2007}, or through mixing  quantum and classical light for polarization entangled states~\cite{NOONmixing1,NOONmixing2}.  Today's state of the art consists of $N=5$ photon N00N states~\cite{Afek879} with most demonstrations using  polarization encoding protocols~\cite{NOONpolar,phase1}. Indeed, while path-encoding offers great potential, it  requires a phase control that is challenging to implement with bulk optics. Recent integrated photonics architectures have enabled the generation of on-chip path-entanglement~\cite{peruzzo2012,brienpath2015,Walther2015,sciarrino-tritter},  thus benefiting from robust and  precise phase control and reconfigurability~\cite{osellame2013}. However, the quantum tomography of multi-photon path-entangled states has been scarcely addressed so far. The tomography of  a path-entangled single photon  can be achieved using quantum homodyne tomography~\cite{babichev2004path} and entanglement witnesses have been derived for two paths~\cite{morin2013path,ho2014path} and were recently extended to multiple paths~\cite{thew2015}. Path-entanglement of two  photons has  been demonstrated on chip, making use of  a path-encoded C-NOT gate~\cite{peruzzo2012} or the equally low probability of generating a photon pair in two non-linear crystals~\cite{brienpath2015}. Yet, in both cases the quantum tomography was achieved for \emph{distinguishable} two-photon states and was mostly intended to quantify the chip performance rather than in-depth characterization of the produced state.

The most natural way of obtaining a two-photon path-entangled N00N state is to perform the Hong-Ou-Mandel (HOM) experiment~\cite{hom} with perfectly indistinguishable single  photons: by  impinging  on the two inputs of a balanced beam splitter, they interfere and leave the beam splitter in a maximally-entangled state---a textbook experiment that has been realized with both heralded~\cite{NOONSPDC1,phase1} and on-demand single photon sources~\cite{Bennett2016}.  To date, the creation of a two-photon N00N state has been supported through the observation of the expected phase dependence  for coincidences measured at the output of a Mach-Zehnder interferometer. Yet, to the best of our knowledge, the full tomography of  path-entangled \emph{indistinguishable} photon states has not been performed, even at the level of two photons.

Here, we propose a novel method to  derive the density matrix of indistinguishable two-photon two-path state. We discuss the accuracy of standard tomography techniques and show how an overcomplete set of measurements  enables us to confidently extract all  coherences  that could be otherwise hidden  because of poor statistics. Finally, by exploiting the bosonic nature of photons as proposed by Adamson and coworkers~\cite{steinberg2007}, we extend our approach to asses the contribution of partially distinguishable photons to the density matrix, which brings insight into the cause for non-maximal entanglement.

\noindent{\bf Generation of the two-photon state}

\begin{figure}[t]
\setlength{\abovecaptionskip}{-5pt}
\setlength{\belowcaptionskip}{-2pt}
\begin{center}
\includegraphics[width=1\linewidth,angle=0]{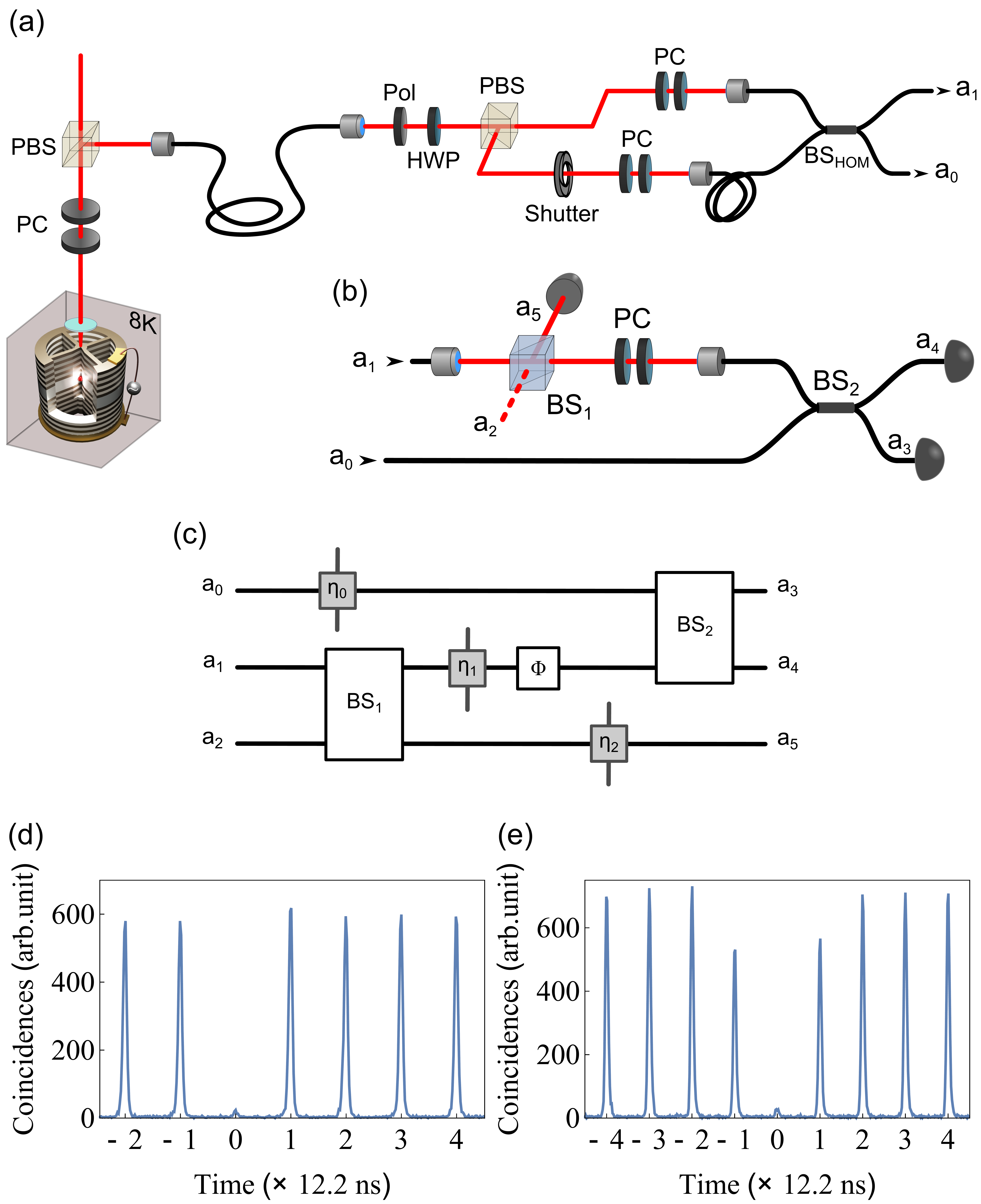}
\end{center}
\caption{  (a) Schematic of the experimental setup used to generate the path entangled two-photon state. HWP stands for half-waveplate, BS for beam splitter,  PBS for polarizing beam splitter and PC for polarization control. The shutter on the lower line is used to measure the phase in the tomography setup every ten seconds. (b) Schematic of the tomography setup. (c) Corresponding mode diagram. (d) Second order correlation function $g^{(2)}(t)$ of the single photon source. (e) Correlation histogram corresponding to $R_{0,1}$. }
\label{Fig1}
\end{figure}

We use a recently developed semiconductor single-photon source \cite{somaschi2016} to generate a two-photon path-entangled state. The device consists of an  electrically-controlled single InGaAs quantum dot (QD) inserted in an optical cavity and placed in a cryostat at 8~K, see Fig.~\ref{Fig1}.a. The  QD exciton transition is resonantly excited  with 15~ps laser pulses at 82~MHz repetition rate. The transition is driven to its excited state  using a $\pi${-}pulse controlled through the laser intensity. The resonant fluorescence photons are collected in a crossed polarization scheme so as to separate them from the excitation laser,  and are subsequently coupled to a single mode optical fiber.  Fig.~\ref{Fig1}(d) shows the coincidence counts obtained when measuring the second order auto-correlation function $g^{(2)}(t)$  with two single photon detectors at the outputs of a fibered beam splitter. The very small area of the peak at zero delay gives $g^{(2)}(0)=0.03\pm0.01$, evidencing the excellent single-photon purity of the source. Note that this residual signal  arises mostly from scattered laser light since no spectral filtering was used in contrast to Ref.~\onlinecite{somaschi2016}. To create the two-photon path-entangled state in a HOM configuration, two photons successively generated $12.2$~ns apart are first  probabilistically routed on both outputs of a free space polarizing beam splitter (PBS), see Fig.~\ref{Fig1}.a. A $12.2$~ns fibered delay line is added to one of the arm  in order to temporally overlap both photons on the fibered HOM beam splitter (BS$_\text{HOM}$), which provides an excellent spatial-mode overlap of 0.997 and well balanced reflection and transmission coefficients R=0.508 and T=0.492. For perfectly indistinguishable photons, the two-photon should exit the beam splitter in the maximally entangled two-photon state $\ket{\psi_{2002}}=\frac{1}{\sqrt{2}}(\ket{2,0}-\ket{0,2})$ where the first (second) number refers to the photon number in the path
$0$ (resp. $1$), see Fig.~\ref{Fig1}.a. Directing the signal of the two output path modes $0$ and $1$ towards single-photon detectors  leads to the standard experimental configuration used to measure the mean wavepacket overlap of the two photons. The corresponding coincidence histogram is shown in Fig.~\ref{Fig1}.e from which a HOM visibility of 0.945 is deduced, corresponding to a mean wavepacket overlap of 0.975 when correcting for the imperfect $g^{(2)}(0)$.

\noindent{\bf Two-photon two-path state quantum tomography}

The state of two photons distributed over two paths, where the two photons cannot be distinguished  in any  degrees-of-freedom  other than their spatial mode,  is described by a $3\times3$ density matrix $\rho^{in}$  in the  $\left|2,0\right>$, $\left|1,1\right>$, $\left|0,2\right>$ basis~\cite{steinberg2007}. 

Tomographical reconstruction of N00N states has been addressed  for two orthogonal polarization modes of one spatial mode \cite{Schilling,israelPRA}, where all coherences can be derived using N-fold coincidences and SU(2) transformations via phase retarders and wave plates. Such scheme can in principle be transposed to path encoding, yet at the cost of stabilizing two independent optical phases: one phase in  one path,  and the other in  an additional Mach-Zender interferometer needed to mimic a tunable beam-splitter. Here, we propose an alternative approach based on a single phase and an ancillary spatial mode.


Fig.~\ref{Fig1}.b  presents the proposed experimental setup and Fig.~\ref{Fig1}.c the corresponding mode diagram. Photons in path $0$ and  $1$, corresponding to the creation operators $\hat{a}^\dagger_0$ and $\hat{a}^\dagger_1$ are sent to a final fibered beam splitter labelled BS2 in a Mach-Zehnder configuration. Path $0$ is directly coupled to one of the inputs of BS2. A free space beam splitter  BS1 is inserted on the other arm of the Mach-Zehnder to entangle path $1$ with the ancillary mode, the path labelled $2$.
The free space part between BS1 and BS2 is not optically stabilized, generating  a slowly varying optical phase $\phi$ which is periodically measured. As shown below, a set of 9 photon correlations measurements, from a proper combination of paths $i$ and $j$, and for two different phases, rendering the correlation rates $R_{i,j}(\phi)$ for $0\le i,j \le 5$,
  allows  performing the quantum state tomography in the spatial mode basis.

This design derives from an  analogy to the tomography of a polarization-entangled  two-photon state \cite{james_measurement_2001} for which a minimal set of measurements---\textit{i.e.} enabling the linear reconstruction of the density matrix---includes photon-correlations between non-orthogonal polarizations. Mapping path $0$ and $1$ to the polarization modes $H$ and $V$, the above experimental configuration essentially mimics such correlation measurements. Detection on the output paths $3$ and $4$   accounts for the projection onto the $(H\pm e^{i\phi}V)/\sqrt{2}$ polarizations. 
Correlations such as $R_{3,4}$---without the additional BS1---evidence a $\cos 2\phi $ dependence, and  have previously been used to confirm the  nature of a two-photon N00N state~\cite{Bennett2016}. Yet a complete polarization tomography must also include  correlations such as $R_{5,i}$ with $i=3,4$, which in the polarization analogy correspond to correlation between linearly polarized photon $V$ and diagonal  or circular polarizations.

To  derive the density matrix from a complete set of measurements, we first consider  the case of a pure input state
 \begin{eqnarray}
 \ket{\psi}^{in}&=&\alpha \ket{2,0}+ \beta \ket{1,1}+\gamma \ket{0,2} \label{Eq1:ket}\\
 &=& (\frac{\alpha}{\sqrt{2}} \hat{a}_{0}^\dagger  \hat{a}_{0}^\dagger +  \beta \hat{a}_{0}^\dagger  \hat{a}_{1}^\dagger+\frac{\gamma}{\sqrt{2}} \hat{a}_{1}^\dagger  \hat{a}_{1}^\dagger) \ket{0_{0},0_{1}}
 \end{eqnarray}

\noindent
and the corresponding density matrix $\rho^{in} $.
The diagonal terms corresponding to the populations can be obtained from the correlation rates  $R_{0,1}$ for  $\ket{1,1}$, $R_{0,0}$ for  $\ket{2,0}$ and $R_{1,1}$ for  $\ket{0,2}$ where $R_{i,j}=\bra{\psi_{in}} \hat{a}_{i}^\dagger\hat{a}_{j}^\dagger \hat{a}_{j}\hat{a}_{i}\ket{\psi_{in}}$ refers to correlation counts obtained by coupling path $i$ and $j$ to detectors. The auto-correlation rate $R_{i,i}$ are obtained by coupling the path $i$ to a beam-splitter and two detectors.  The population of the $\left|1,1\right>$  state ranges from $0$, in case of perfectly indistinguishable photons, to $0.5$ for fully distinguishable ones.
By making use of the unitary transformation between modes 0,1,2 and modes 3,4,5---determined by the optical setup  $$\begin{pmatrix}
\hat{a}_{3}^\dagger \\
\hat{a}_{4}^\dagger \\
\hat{a}_{5}^\dagger
\end{pmatrix}=   U_{setup} \begin{pmatrix}
\hat{a}_{0}^\dagger \\
\hat{a}_{1}^\dagger \\
\hat{a}_{2}^\dagger
\end{pmatrix}$$


\noindent  we  calculate the output state $\ket{\psi_{out}}$ and the corresponding correlation rates  $R_{i,j}=\bra{\psi_{out}} \hat{a}_{i}^\dagger\hat{a}_{j}^\dagger \hat{a}_{j}\hat{a}_{i}\ket{\psi_{out}}$ as a function of the density matrix elements $\rho^{in}_{k,l}$.  By doing so, a minimal set of correlation measurements $R_{\text{ comp.}} (\phi_1,\phi_2)$ is obtained when measuring the following rates for two distinct phases $\phi_1$ and $\phi_2$:

\begin{align*}
R_{\text{ comp.}} (\phi_1,\phi_2)&=  \bigg(R_{0,0}\ ,R_{0,1}\ ,R_{1,1}\ ,R_{3,3}(\phi_1)\ ,R_{3,4}(\phi_1)\ ,\\
& R_{4,5}(\phi_1)\ ,R_{3,3}(\phi_2)\ ,R_{3,4}(\phi_2)\ ,R_{4,5}(\phi_2)\bigg)
\end{align*}
 with  $\vert \phi_1-  \phi_2\vert  \neq 0, \frac{\pi}{2}, \pi$. The corresponding linear transformation matrix $M$ relating $R_{\text{ comp.}}$ to the vectorial form of $(\rho^{in})$ is invertible so that the density matrix of the analyzed state is deduced from correlation measurements through the linear equation: \begin{equation} (\rho^{in})=M^{-1}R_{\text{ comp.}} (\phi_1,\phi_2) \end{equation}
 \noindent

 \noindent The same relation holds for any mixed input state for which the density matrix  is a linear superposition of  pure-state density matrices  weighted by the corresponding state probability.

In practice,  some optical losses on the setup, related to fiber to fiber, or free-space to fiber coupling, should be considered to model the corresponding correlations.
These losses are modeled as additional beam splitters, labelled $\eta_{i}$ for $i=0,1,2$ as shown in Fig.~\ref{Fig1}.c.
Such approach allows maintaining a unitary description of the experiment and keeping the  same procedure as described above, at the cost of introducing more modes.
\begin{figure}[t]
\setlength{\abovecaptionskip}{-5pt}
\setlength{\belowcaptionskip}{-2pt}
\begin{center}
\includegraphics[width=1\linewidth,angle=0]{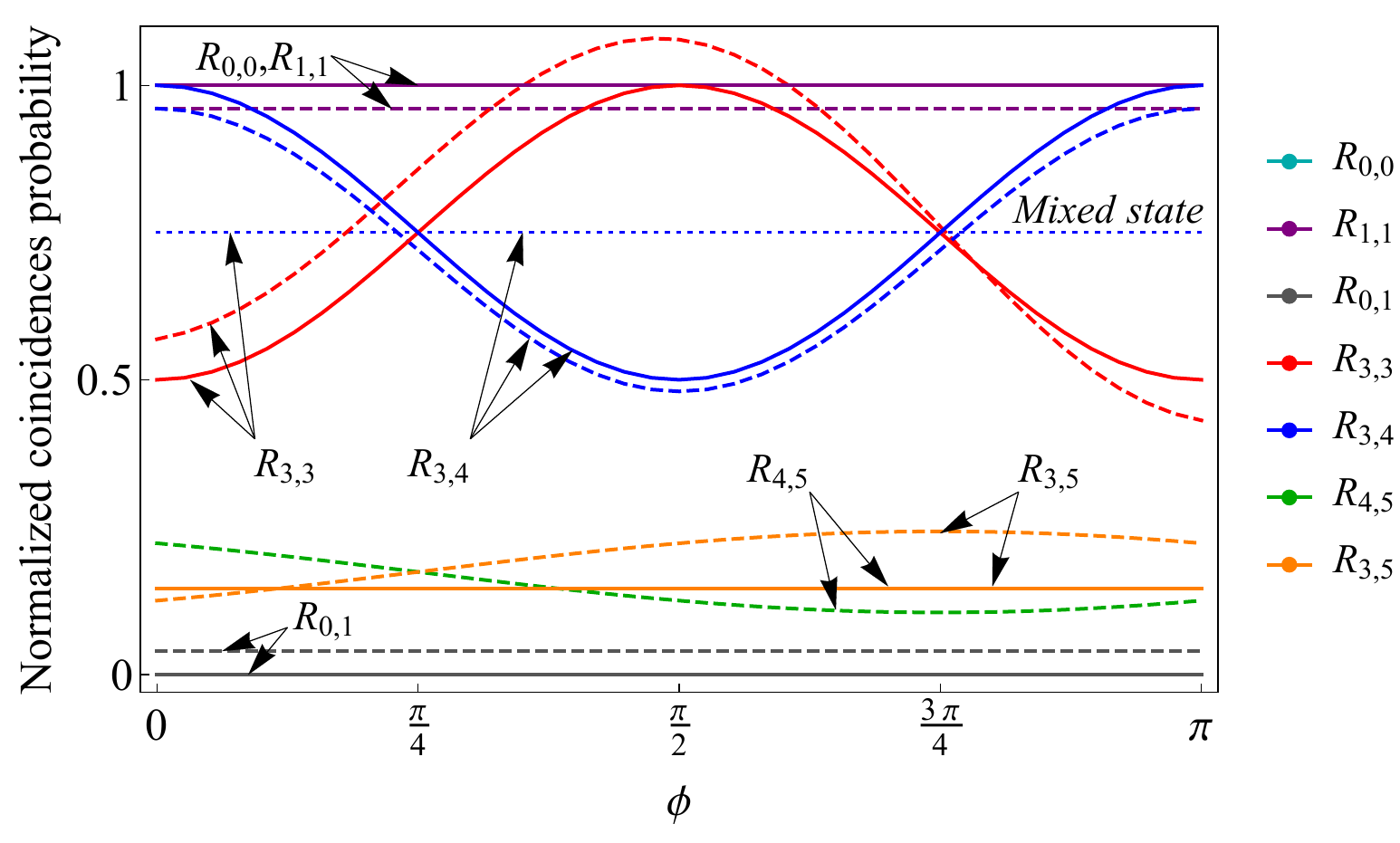}
\end{center}
\caption{ Symbols: measured correlation rate $R_{i,j}$ as a function of the phase $\phi$. Solide lines: calculated correlation rates for the ideal two-photon state $\frac{1}{\sqrt{2}}(\ket{2,0}+\ket{0,2})$. Dotted lines: calculated correlation rates for a mixture of $\ket{2,0}$ and $\ket{0,2}$ states. In this case $R_{3,4}=R_{3,3}$ is independent on $\phi$. Dashed lines:  calculated correlation rates corresponding to the state $\ket{\psi}^{in}=\frac{\cos(\theta)}{\sqrt{2}} \ket{2,0}{+}\sin(\theta)  e^{i\frac{\pi}{4}}\ket{1,1}{-}\frac{\cos(\theta)}{\sqrt{2}}  \ket{0,2}$, with $\theta{=}0.2$\label{Fig2}}
\end{figure}

The calculated coincidence rates are   shown Fig.~\ref{Fig2}  for various input states in order to illustrate the sensitivity of the corresponding measurements.  
In the case of the ideal maximally entangled state  $\ket{\psi_{2002}}$ (solid lines), both coincidence count rates $R_{3,4}$ and $R_{3,3}$ are expected to vary with $\phi$, with a  maximum contrast being determined by the coherence term between $\ket{0,2}$ and $\ket{2,0}$ and the losses in the Mach-Zehnder. For a fully mixed two-photon state (dotted lines), all coincidences  show no dependence on $\phi$ (overlapping dotted red and blue lines for $R_{3,4}$ and $R_{3,3}$). The dashed line shows the calculated rates for the pure state $\ket{\psi}^{in}=\frac{\cos(\theta)}{\sqrt{2}} \ket{2,0}{+}\sin(\theta)  e^{i\frac{\pi}{4}}\ket{1,1}{-}\frac{\cos(\theta)}{\sqrt{2}}  \ket{0,2}$, with $\theta{=}0.2$. The population on the $ \ket{1,1}$ component results in a $\cos \phi$ dependence of $R_{3,5}$ and $R_{4,5}$, shifted by its initial phase---$\pi/4$ in the present example---as a result of interference  in the Mach-Zehnder which produces a dephasing $\pm\phi$  of the  ket $\ket{1,1}$  with respect to  $\ket{0,2}$ and $\ket{2,0}$. The corresponding coherences also imprint a  $\phi$ dependence on top of the $2\phi$ modulation in $R_{3,3 }$, responsible for a  small asymmetry.

\vspace{0,2cm}
\noindent{\bf Overcomplete set of measurements}
\vspace{0,2cm}

\begin{figure}[h]
\setlength{\abovecaptionskip}{-5pt}
\setlength{\belowcaptionskip}{-2pt}
\begin{center}
\includegraphics[width=0.95\linewidth,angle=0]{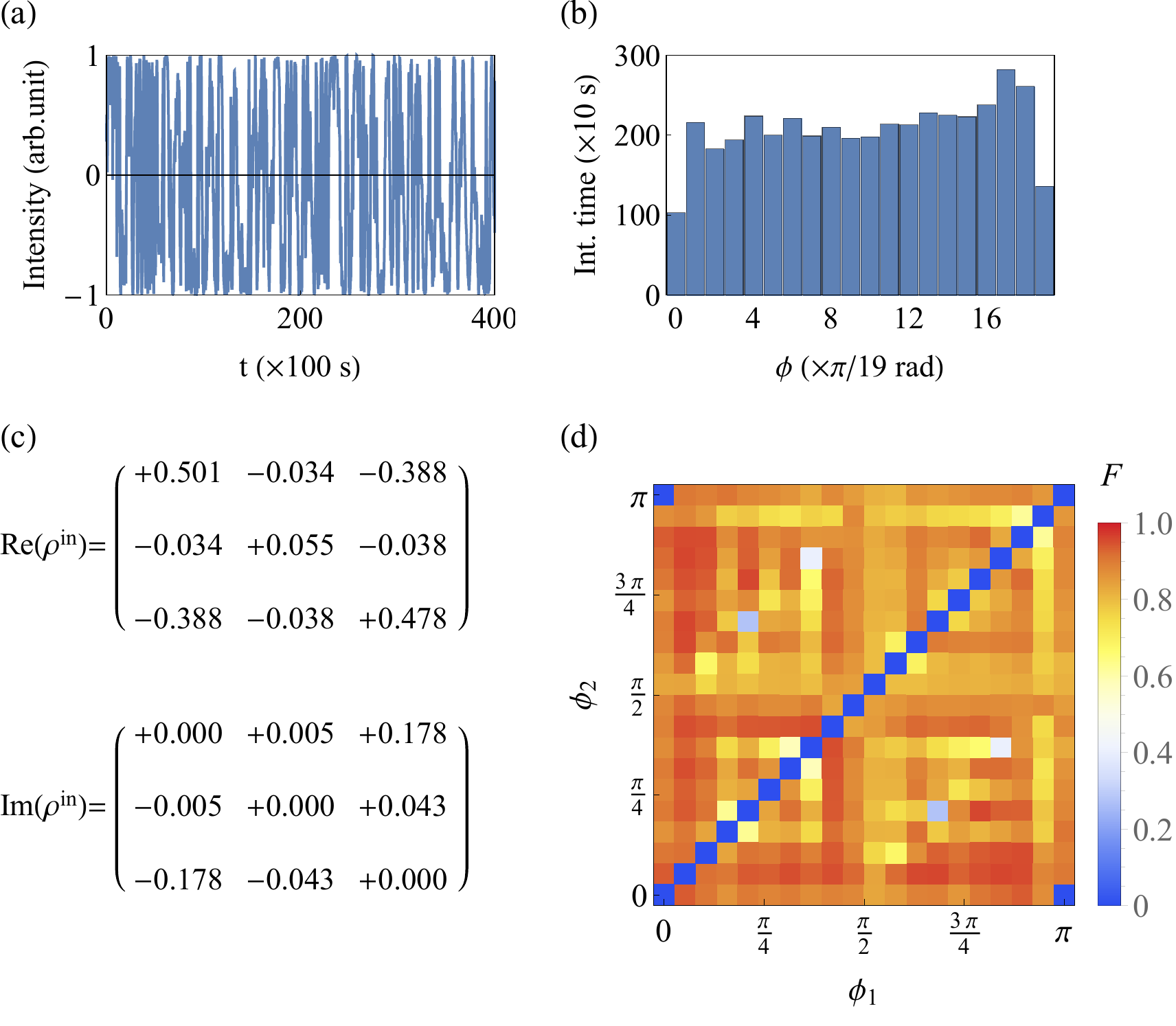}
\end{center}
\caption{(a): Single photon rate giving access the the phase $\phi$ in the tomography setup, measured  as a function of time on path 3 when blocking the lower path in Fig. \ref{Fig1}.a.  (b) Corresponding histogram of the acquisition time periods as a function of  20 phase bins. (c) Real and imaginary parts of the density matrix deduced from a set of nine measurements using the linear inversion tomography for $(\phi_1,\phi_2)=(0,\frac{\pi}{4})$. (d) Fidelity to the maximally entangled N00N state deduced from the maximum  likelihood method with 9 measurements as a function of  $(\phi_1,\phi_2)$.  }
\label{Fig3}
\end{figure}

To obtain the correlation rates $R_{3,4}$, $R_{3,5}$ and $R_{4,5}$, we use the experimental configuration sketched in  Fig.~\ref{Fig1}.b. The phase $\phi$ in the interferometer arm freely evolves over time and correlation counts are continuously acquired. The phase is measured every ten seconds by closing one input path of the HOM beam splitter $BS_{HOM}$ using an electronically controlled shutter so that only one-photon path enters the analysis setup.
The intensity signal recorded on path $3$ or $4$ is due to the single photon interference, and oscillates with $\phi$ giving access to its time dependence.
Fig.~\ref{Fig3}.a shows the time trace of the corresponding signal  recorded over a ten-hour period.
It shows large fluctuations of $\cos(\phi)$  indicating that $2\pi$ variations of $\phi$  take place over a typical 10~min timescale. Fig.~\ref{Fig3}.b shows the corresponding histogram of total acquisition times distributed over 20 phase bins, showing a reasonably flat dependence with $\phi$.

Three detectors are used on path 3,~4 and 5 to record the three detection counts simultaneously. Time tagging  of the events on the three detectors is recorded with respect to the laser trigger in order to reconstruct the correlation rates as a function of $\phi$. To remove the errors due to fluctuations of the signal over time---arising from mechanical fluctuations in the relative laser spot-source overlap---the coincidences counts of each measurement is normalized. Normalization is achieved with the correlation peaks recorded at time delays corresponding to multiple of the laser repetition period ($k \times 12$~ns)  with $\vert k\vert \ge 2$.
These  peaks  are due to  single photon events arising from different excitation pulses and their magnitude can also be  theoretically predicted from  the product of single detection rates $R_{j}=\bra{\psi_{out}} \hat{a}_{j}^\dagger \hat{a}_{j}\ket{\psi_{out}}$.

\vspace{0,2cm}
\noindent {\bf Standard quantum tomography }
\vspace{0,2cm}

We first use the standard linear tomography approach, making use of Eq.3. As discussed by Thew and co-authors, linear quantum tomography does not require that the projectors forming a complete set of measurements are orthogonal \cite{Thew2002}. Mathematically,  any couple of phases such that $\vert\phi_1 - \phi_2\vert \neq 0,\frac{\pi}{2},\pi$, $R_{\text{ comp.}} (\phi_1,\phi_2)$ allows a reconstruction of the density matrix. 
 Indeed, we note  that $(\phi_1,\phi_2)=(\frac{\pi}{4},\frac{3\pi}{4})$  does not allow discriminating between the ideal  N00N state from the totally mixed state, see solid and dotted lines in Fig.~\ref{Fig2}.c. In this case, $R_{34}$ and $R_{44}$  turn out be only sensitive to the imaginary part of the coherence. 
In order to determine with the same precision the real and imaginary parts of all coherences an optimal choice providing the lowest uncertainties in the state tomography is found  for a couple of phases such that $\vert\phi_2-\phi_1\vert\approx\frac{\pi}{4}$.

As an example, we  derive  the raw density matrix obtained for $(\phi_1,\phi_2)=(0,\frac{\pi}{4})$ in Fig.\ref{Fig3}.c. It exhibits small deviations from a physical density matrix with notably $\text{Tr}(\rho^{in})=1.034>1$. To determine a meaningful value of fidelity, we  normalize the unphysical state by the trace and obtain a   fidelity to the ideal $\ket{\psi_{2002}}$ state of $F=0.85$.
 To avoid the issue of non-physical properties of the resulting matrix~\cite{james_measurement_2001}, we use the maximum likelihood approach  and numerically determine  nine parameters $t_{\nu}$ defining the physical density matrix $\rho^{in}(t_1,..t_9)$, to maximize the likelihood function $$P(t_1,..t_9)=\prod_{\nu=1}^{9}\exp {-\frac{R_\nu(t_1,..,t_9)-R_\nu}{\sigma_\nu^2}}$$ where $R_\nu(t_1,..,t_9)$ are the expected coincidence rate for the test input state $\rho^{in}(t_1,..t_9)$,  $R_\nu$ are the measured ones. $\sigma_\nu$ is the standard deviation of the $\nu^{th}$ coincidence.

 Fig.~\ref{Fig3}.d. shows the fidelity to  $\rho_{2002}$ deduced using the maximum likelihood method as a function of $(\phi_1,\phi_2)$. Fluctuations in the  pattern of  Fig.~\ref{Fig3}.d  result from experimental noise and/or insufficient statistics. The fidelity  drops in the vicinity of $(\phi_1,\phi_2)=(\pi/4,3\pi/4)$ as expected from the  discussion above.  We have calculated the average value of the deduced fidelity  as well as its standard deviation as a function of $\vert\phi_2-\phi_1\vert$: no dependence is observed for the present set of measurements as long as $\vert\phi_2-\phi_1\vert\neq0 $ and $\vert\phi_2-\phi_1\vert \neq \frac{\pi}{2} $ where noise gives a strong threefold increase of the standard deviation. Away from these two singular points, the average fidelity is $0.87$ with a standard deviation of $0.06$.

\begin{figure}[t]
\setlength{\abovecaptionskip}{-5pt}
\setlength{\belowcaptionskip}{-2pt}
\begin{center}
\includegraphics[width=1\linewidth,angle=0]{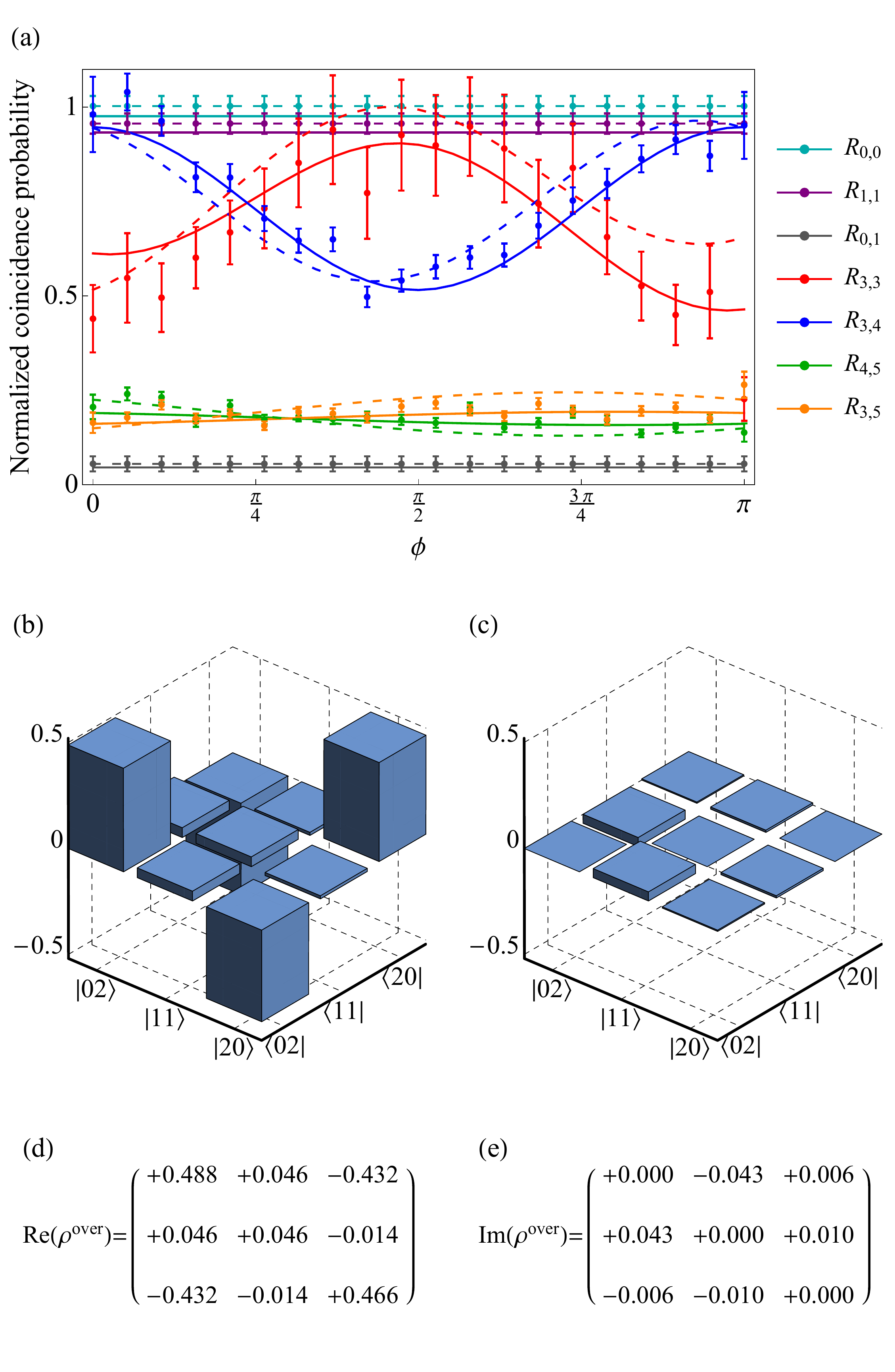}
\end{center}
\caption{ a: Normalized coincidence rates as a function of $\phi$. Symbols: measurements. Dotted lines: calculated coincidence rates for the state deduced from linear tomography for $(\phi_1,\phi_2)=(0,\frac{\pi}{4})$ (see Fig.\ref{Fig3}.b). Solid lines: Coincidence rates calculated for the state deduced from the vercomplete set of 79  measurements. (b-d) Real  and (c-e) imaginary part of the density matrix deduced from an overcomplete set of 79 measurements. }
\label{Fig4}
\end{figure}

\vspace{0,2cm}

\noindent{\bf Overcomplete quantum tomography}
\vspace{0,2cm}

The measured normalized coincidence count rates $R_{3,4}$, $R_{3,3}$, $R_{3,5}$ and $R_{4,5}$ are plotted  in Fig.~\ref{Fig4}.a as a function of $\phi$, together with the phase-independent count rates $R_{0,0}$, $R_{0,1}$ and $R_{1,1}$.  Experimental error bars are derived taking into account  the Poissonian noise on the  coincidences as well as on the normalization count rates.
 $R_{3,3}$ shows stronger noise due to the lower statistics available for the measurement.

The dotted lines in Fig.~\ref{Fig4}.a show the  correlation rates calculated for the state deduced from linear tomography presented in Fig.\ref{Fig3}.c., evidencing the limited accuracy of the standard tomography: the corresponding correlations   fail to reproduce the experimental ones on the full $\phi$ scale.

 To obtain a   better insight into the two-photon state, the likelihood function is now maximized for the whole set of 79 phase dependent measurements. The density matrix of the corresponding input state $\rho_{\text{over}}$ is shown in  Fig.~\ref{Fig4}.b-e. It presents a fidelity to the N00N state of $0.91$,  in the range of the average fidelity  obtained through 9 measurements within the standard deviation. 
The coincidence rates corresponding to this reconstructed  state is superimposed to the measurements in  Fig.~\ref{Fig4}.a (solid lines). It shows a very good agreement with the experimental observations.  The observed small phase dependence of $R_{4,5}$ and $R_{3,5}$ is  well accounted for, evidencing coherence between the $\ket{1,1}$ and the  $\ket{0,2}$ and $\ket{2,0}$ terms. This analysis shows the reliabillity of the information that can be extracted from such an overcomplete data set. In the next section, we extend our analysis a step further to obtain a diagnosis for the deviation of the produced state from the ideal N00N state.

\vspace{0,2cm}

\noindent{\bf Extracting  the true photon indistinguishability}
\vspace{0,2cm}

The creation of a maximally entangled 2-photon N00N state depends on various parameters: the indistinguishability of the photons impinging on the HOM beam splitter, the balance of the reflection and transmission  coefficients, as well as any undesired source of background light. On one hand, the interference of two perfectly indistinguishable photons on an unbalanced beam splitter, with $|R|\neq |T|$, results in a $\ket{1,1}$ population. On the other hand,  two distinguishable photons create a  $\ket{1,1}$ population with a perfectly balanced beam splitter.  In any case, the distinguishability of the photons affects  coherences between $\ket{2,0}$ and $\ket{0,2}$ only via the reduction of their populations. In the present experiment---using a semiconductor quantum dot operated without any spectral filtering of the zero-phonon line---two origins for the photon distinguishability can be expected. First, a residual phonon sideband emission certainly takes place and slightly reduces the photon indistinguishability, as recently shown~\cite{Thomas2017}. Additionally,  the resonant excitation scheme  leads to a small fraction of residual laser light   not completely suppressed in the crossed polarized collection. This residual light is also distinguishable from the single photons emitted by the quantum dot and is also responsible for the measured non-zero $g^{(2)}(0)$   shown Fig.~\ref{Fig1}.d.

\begin{figure}[t]
\setlength{\abovecaptionskip}{-5pt}
\setlength{\belowcaptionskip}{-2pt}
\begin{center}
\includegraphics[width=1\linewidth,angle=0]{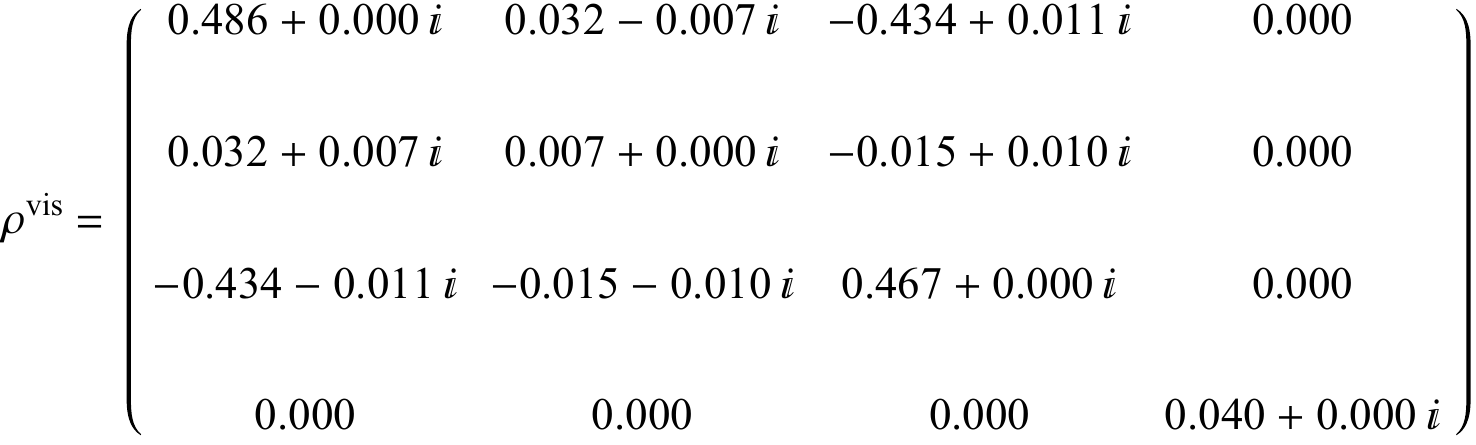}
\end{center}
\caption{ 4x4 density matrix with the distinguishable and indistinguishable part of the two-photon state. }
\label{Fig5}
\end{figure}

Even though the physical origin of each detected photon cannot be determined by our apparatus used for tomography, Adamson and coworkers have demonstrated that in such situation  it is still possible to get more information on the two-photon state~\cite{steinberg2007}.  The  contribution of the  truly distinguishable photons to the $\ket{1,1}$ population can  be separated from that due to an imperfect set up via a more refined analysis. 
In practice, one  introduces the 4-state basis   $\ket{2,0}$, $\ket{0,2}$,  $\psi^+$, $\psi^-$, corresponding to the visible degree of freedom, where the $\psi^{\pm}$ are now the symmetric and antisymmetric states of  two possibly distinguishable  photons "a" and "b"  on each path:
$\psi^{\pm}= \frac{\ket{1_a,1_b}\pm\ket{1_b,1_a}}{\sqrt{2}}$. Truly indistinguishable photons can only occupy the symmetric state $\psi^{+}$, thus any population in the antisymmetric state $\psi^{-}$ reveals the presence of distinguishing information.
The 4x4 density matrix $\rho^\text{vis}$  reads in this basis~\cite{steinberg2007,NOONSPDC1}:

\begin{eqnarray}
\rho^\text{vis} & =&
\begin{pmatrix}
\rho_{20,20} & \rho_{20,\psi^+} & \rho_{20,02} & 0\\
\rho_{\psi^+,02} & \rho_{\psi^+,\psi^+}& \rho_{02,\psi^+} & 0\\
\rho_{02,20} & \rho_{\psi^+,02} & \rho_{02,02} & 0\\
0 & 0 &0 &\rho_{\psi^-,\psi^-}
\end{pmatrix}\\
& = & \begin{pmatrix}
\Bigg(\rho^{+}_{k,l}\Bigg) & 0 \\
0 & \rho^{-}
\end{pmatrix}
\end{eqnarray}

\noindent where the coherences between the 3x3 symmetric $\rho^{+}$ and 1x1 antisymmetric $\rho^{-}$ subspaces  are zero.
By considering a pure input state in the form $$
\ket{\psi}^{in}= \bigg(\alpha \hat{a}_{0,a}^\dagger  \hat{a}_{0,b}^\dagger +  \beta \hat{a}_{0,a}^\dagger  \hat{a}_{1,b}^\dagger+\gamma \hat{a}_{0,b}^\dagger  \hat{a}_{1,a}^\dagger+\delta \hat{a}_{1,a}^\dagger  \hat{a}_{1,b}^\dagger \bigg) \ket{0}
$$ and calculating the corresponding coincidences  $$R_{i,j}=\bra{\psi_{out}} \hat{a}_{i,a}^\dagger\hat{a}_{j,b}^\dagger \hat{a}_{j,b}\hat{a}_{i,a}+\hat{a}_{i,b}^\dagger\hat{a}_{j,a}^\dagger \hat{a}_{j,a}\hat{a}_{i,b}\ket{\psi_{out}},$$  we can determine new relations between $(\rho^{+}_{k,l},  \rho^{-})$ and the  $R_{i,j}$ terms. We observed that, as expected, the calculated $R_{i,j}$  do not formally depend on the coherence terms   between the symmetric and antisymmetric part of the density matrix, even if they are considered as non-zero. We then carry out the maximum likelihood  method using the overcomplete set of measurements to obtain the ten parameters defining the physical  density matrix in the form of  $\rho^\text{vis}$, see Fig.~\ref{Fig5}. Notably, most of the $\ket{1,1}$ population now appears on the antisymmetric part $\rho_{\psi^-,\psi^-}$ of the density matrix, with a negligible population on the symmetric $\rho_{\psi^+,\psi^+}$ population. This approach allows us to ascribe most of  the N00N  state imperfection to a partial distinguishability of the photons and not to imperfections in the HOM beam-splitter. Furthermore, knowing that the lower bound of $\rho_{\psi^+,\psi^+}+\rho_{\psi^-,\psi^-}$  is given by $g^{(2)}(0)=0.03$, we  ascribe most of the extracted  distinguishability to the residual laser.


{\bf Conclusions}

In the present work, we have proposed a simple experimental method to perform  quantum state tomography of a two-photon path-entangled state. Although unavoidable experimental noise lead to uncertainties in a standard quantum tomography approach, we have shown that an overcomplete data set allows extracting  highly reliable information. Moreover, accessing  the indistinguishable and distinguishable parts of the density matrix, we can provide   a precise diagnosis as for the deviation from the ideal state, separating the limitations arising from the photon source to those coming from the imperfections of the optical network.


High-photon number path-entangled N00N states are foreseen as  important resources for many applications ranging from quantum imaging to  quantum sensing and lithography and yet, the possibility to universally detect entanglement without performing a full state tomography is still debated~\cite{FST}. The quantum  tomography of  polarization-encoded N00N state has been extended to high $N$ by making use of photon-number resolving detectors~\cite{Schilling,israelPRA}. Applying a similar extension to access all the required $N^\text{th}$ order photon correlations, we expect that our approach  offers a viable method for the quantum tomography of path-encoded N00N states for any $N>2$.


{\bf Acknowledgements:} The authors are thankful to Pr. Andrew G. White  for  stimulating discussions. This work was partially supported by the  ERC Starting Grant No. 277885 QD-CQED, the French Agence Nationale pour la Recherche grant USSEPP, the French RENATECH network, the Marie Curie individual fellowship SQUAPH, a public grant overseen by the French National Research Agency (ANR) as part of the "Investissements d'Avenir" program (Labex NanoSaclay, reference: ANR-10-LABX-0035), and the iXcore foundation.


\end{document}